\documentclass[pra,twocolumn,aps,floatfix,showpacs,tightenlines,superscriptaddress,amsmath,amssymb,showkeys,10pt,nofootinbib]{revtex4-2}
\usepackage{slashed}
\usepackage{xcolor}
\usepackage{graphicx}
\usepackage{hyperref}
\usepackage{placeins}
\usepackage{dsfont}
\usepackage{amsthm}

\newcommand{\be}{\begin{equation}}\newcommand{\ee}{\end{equation}}
\newcommand{\bea}{\begin{eqnarray}}\newcommand{\eea}{\end{eqnarray}}
\newcommand{\brr}{\begin{array}}\newcommand{\err}{\end{array}}
\newcommand{\bit}{\begin{itemize}}\newcommand{\eit}{\end{itemize}}
\newcommand{\ben}{\begin{enumerate}}\newcommand{\een}{\end{enumerate}}

\newcommand{\bbm}{\begin{bmatrix}}\newcommand{\ebm}{\end{bmatrix}}
\newcommand{\ba}{\begin{array}}
\newcommand{\ea}{\end{array}}

\newtheorem{mydef}{Definition}
\newtheorem{Lemma}{Lemma}
\newcommand{\bd}{\begin{mydef}} \newcommand{\ed}{\end{mydef}}
\newcommand{\bthe}{\begin{theorem}} \newcommand{\ethe}{\end{theorem}}
\newcommand{\ble}{\begin{Lemma}} \newcommand{\ele}{\end{Lemma}}

\def\lan{\langle}
\def\lf{\left}

\def\non{\nonumber}\def\ran{\rangle}

\def\ri{\right}
\def\Ga{\Gamma}

\def\1{{_{1}}}\def\2{{_{2}}}

\def\noHe0{:\;\!\!\;\!\!:H_e(0):\;\!\!\;\!\!:}
\def\noHm0{:\;\!\!\;\!\!:H_\mu(0):\;\!\!\;\!\!:}

\def\lan{\langle}
\def\lf{\left}

\def\non{\nonumber}
\def\ran{\rangle}

\def\ri{\right}

\def\Ga{\Gamma}

\def\1{{_{1}}}\def\2{{_{2}}}

\makeatletter
\let\save@mathaccent\mathaccent
\newcommand*\if@single[3]{%
  \setbox0\hbox{${\mathaccent"0362{#1}}^H$}%
  \setbox2\hbox{${\mathaccent"0362{\kern0pt#1}}^H$}%
  \ifdim\ht0=\ht2 #3\else #2\fi
  }
\newcommand*\rel@kern[1]{\kern#1\dimexpr\macc@kerna}
\newcommand*\widebar[1]{\@ifnextchar^{{\wide@bar{#1}{0}}}{\wide@bar{#1}{1}}}
\newcommand*\wide@bar[2]{\if@single{#1}{\wide@bar@{#1}{#2}{1}}{\wide@bar@{#1}{#2}{2}}}
\newcommand*\wide@bar@[3]{%
  \begingroup
  \def\mathaccent##1##2{%
    \let\mathaccent\save@mathaccent
    \if#32 \let\macc@nucleus\first@char \fi
    \setbox\z@\hbox{$\macc@style{\macc@nucleus}_{}$}%
    \setbox\tw@\hbox{$\macc@style{\macc@nucleus}{}_{}$}%
    \dimen@\wd\tw@
    \advance\dimen@-\wd\z@
    \divide\dimen@ 3
    \@tempdima\wd\tw@
    \advance\@tempdima-\scriptspace
    \divide\@tempdima 10
    \advance\dimen@-\@tempdima
    \ifdim\dimen@>\z@ \dimen@0pt\fi
    \rel@kern{0.6}\kern-\dimen@
    \if#31
      \overline{\rel@kern{-0.6}\kern\dimen@\macc@nucleus\rel@kern{0.4}\kern\dimen@}%
      \advance\dimen@0.4\dimexpr\macc@kerna
      \let\final@kern#2%
      \ifdim\dimen@<\z@ \let\final@kern1\fi
      \if\final@kern1 \kern-\dimen@\fi
    \else
      \overline{\rel@kern{-0.6}\kern\dimen@#1}%
    \fi
  }%
  \macc@depth\@ne
  \let\math@bgroup\@empty \let\math@egroup\macc@set@skewchar
  \mathsurround\z@ \frozen@everymath{\mathgroup\macc@group\relax}%
  \macc@set@skewchar\relax
  \let\mathaccentV\macc@nested@a
  \if#31
    \macc@nested@a\relax111{#1}%
  \else
    \def\gobble@till@marker##1\endmarker{}%
    \futurelet\first@char\gobble@till@marker#1\endmarker
    \ifcat\noexpand\first@char A\else
      \def\first@char{}%
    \fi
    \macc@nested@a\relax111{\first@char}%
  \fi
  \endgroup
}
\makeatother
\begin{document}

\title{Tests of macrorealism in meson oscillation physics}

\author{Massimo Blasone}
\email{blasone@sa.infn.it}
\affiliation{Dipartimento di Fisica, Universit\`a di Salerno, Via Giovanni Paolo II 132, 84084 Fisciano (SA), Italy}
\affiliation{INFN Sezione di Napoli, Gruppo collegato di Salerno, Italy}
\author{Fabrizio Illuminati}
\email{filluminati@unisa.it}
\affiliation{INFN Sezione di Napoli, Gruppo collegato di Salerno, Italy}
\affiliation{Dipartimento di Ingegneria Industriale, Universit\`a di Salerno,
Via Giovanni Paolo II 132, 84084 Fisciano (SA), Italy}
\author{Luciano Petruzziello}
\email{lupetruzziello@unisa.it}
\affiliation{INFN Sezione di Napoli, Gruppo collegato di Salerno, Italy}
\affiliation{Dipartimento di Ingegneria Industriale, Universit\`a di Salerno,
Via Giovanni Paolo II 132, 84084 Fisciano (SA), Italy}
\affiliation{Institut f\"ur Theoretische Physik, Albert-Einstein-Allee 11, Universit\"at Ulm, 89069 Ulm, Germany}
\author{Kyrylo Simonov}
\email{kyrylo.simonov@univie.ac.at}
\affiliation{Fakult\"{a}t f\"{u}r Mathematik, Universit\"{a}t Wien, Oskar-Morgenstern-Platz 1, 1090 Vienna, Austria}
\author{Luca Smaldone}
\email{lsmaldone@unisa.it}
\affiliation{Dipartimento di Fisica, Universit\`a di Salerno, Via Giovanni Paolo II 132, 84084 Fisciano (SA), Italy}
\affiliation{INFN Sezione di Napoli, Gruppo collegato di Salerno, Italy}

\begin{abstract}
Macrorealism formalizes the seemingly intuitive notion that, in contrast with the principles of quantum mechanics, a physical system can be in a definite state at any given time and moreover its dynamical evolution is independent of the measurements performed on it. In this study, we carry out a comparative analysis between {three-time} Leggett--Garg-type inequalities and the conditions of no-signaling-in-time and arrow-of-time for macrorealism within the context of meson oscillations. Our findings indicate that, under given initial conditions, no violations of Leggett--Garg inequalities are observed. However, no-signaling-in-time conditions are found to be violated, thereby revealing the impossibility of applying a macrorealistic description to the physics of meson oscillations.
\end{abstract}

\maketitle
\section{Introduction}

The intersection of particle physics, quantum field theory, and quantum gravity with the study of fundamental conceptual quantum structures and quantum foundations stands out as one of the most active and dynamic domains in contemporary physics research \cite{Bertlmann2003, Blasone2008, Blasone2009, Donadi2012, Donadi2013, Bahrami2013, Blasone2014, SimonovLetter2016, Formaggio2016, Gango2017, Simonov2018, Dixit:2018gjc, Capolupo2019, Simonov2019, Naikoo:2019gme, Naikoo:2020LGI,  Simonov2020, Simonov2021, Zhang2021, Petruzziello2021, Santini2022, Halliwell2022, Wang2022, Simonov2022, Blasone:2023mbc, Blasone:2023iwf, Bosso2023, Petruzziello2023, Bosso2024}. A pertinent illustration is given by the phenomenon of flavor mixing and oscillations, such as the case of neutrino oscillations. This phenomenon not only indicates physics beyond the Standard Model, but has provided a suitable experimental platform for testing the validity of quantum mechanics itself. In particular, the concept of macroscopic realism (\textit{macrorealism}), which encodes the intuition suggested by the experience of our everyday macroscopic world and is in conflict with predictions of quantum mechanics, has been put under scrutiny within flavor oscillating systems \cite{Formaggio2016,Wang_2017,Wang2022}. The violation of its formal occurrence in the quantum realm is usually provided by a set of relations known as the \emph{Leggett--Garg inequalities (LGIs)} \cite{PhysRevLett.54.857, PhysRep2014}, which are deemed as the temporal counterparts of Bell inequalities \cite{PhysRevLett.54.857, Brukner2007, Brukner2008, PhysRep2014, Kumari2017}. Specifically, a violation of LGIs gives a benchmark of quantumness of the underlying system as the inability of meeting all the assumptions of macrorealism \cite{PhysRevLett.54.857}.

It is worth stressing, however, that LGIs provide a necessary but not a sufficient condition for macrorealism \cite{Clemente2015, Clemente2016}. 
Nevertheless, it is still possible derive a set of equalities known as \emph{no-signaling-in-time} (NSIT) \cite{Brukner2008, Clemente2015} and \emph{arrow-of-time} (AoT) conditions offering a one-to-one correspondence with macrorealism \cite{Kofler2013, Clemente2015}. Indeed, the quantumness of neutrino oscillations can be unveiled by NSIT/AoT conditions even under fulfillment of LGIs, e.g., when wave-packet decoherence is considered \cite{Blasone:2023iwf}.

In this work, we question the validity of macrorealism via NSIT/AoT conditions in oscillations of neutral mesons. In contrast to neutrinos, neutral mesons are decaying particles, a feature which adds an additional layer of complexity in macrorealistic scenario that demands consideration. Focusing on neutral kaons (the most studied neutral meson systems) and employing an approach grounded in conditional probabilities, we derive LGIs in its canonical form as well as in another reformulation, commonly referred to as LGIs in Wigner form (WLGIs), and compare them with predictions of the NSIT/AoT conditions. Under specific initial conditions, no violations of LGIs or WLGIs are found, while the NSIT/AoT conditions persistently exhibit violations in such instances. This result underscores the robustness and efficacy of the proposed methodology. It is important to stress that the findings presented in this paper can be equally obtained by treating particle decay within the framework of an open quantum system and repeating our computations using the Kraus operators formalism \cite{Caban:2005ue, Caban:2006ij, Castillo2013, Naikoo2018B}.
 
The remainder of the paper is organized as follows: in Section~\ref{sec:2}, we review the notion of macrorealism and the related conditions we will employ in our analysis (i.e. LGIs, WLGIs and NSIT/AoT). Moreover, we give the basics of meson oscillations phenomenology by resorting to the explicit example of neutral kaons. Then, in Section~\ref{sec:3} we analyze the ensuing LGIs, WLGIs and NSIT/AoT conditions; with these results, we then establish a comparison among the various tests of macrorealism. Finally, Section \ref{sec:4} contains conclusions and future perspectives. 

The explicit computation of joint probabilities is reported in Appendix~\ref{app:A} for reader's convenience.
\section{Preliminaries}\label{sec:2}
\subsection{Macrorealism and its conditions}\label{sec:2A}

Quantum mechanics is known as a theory offering a highly accurate description of Nature, as confirmed by numerous experiments. The  fundamental superposition principle manifests itself in a wide variety of phenomena at different energy scales, e.g., particle mixing and oscillations which are experimentally observed in various systems just like neutrinos and neutral mesons. Nevertheless, quantum mechanics exhibits several conceptual issues which are still actively debated. For example, a naive attempt to apply the superposition principle at the macroscopic scale inevitably leads to paradoxes, such as infamous Schr\"odinger's cat. Indeed, our everyday experience suggests a breakdown of the superposition principle and the emergence of the laws of classical physics, thus leading to a natural question: Why and how does the quantum world blend in the everyday classical world? 

Since the ``standard" quantum mechanics does not go beyond postulating an ad hoc separation between them, various interpretations and modifications of the quantum theory (e.g., Bohmian mechanics and spontaneous collapse models) have been developed in order to explain the quantum-to-classical transition. One of the ways to explore it is probing coherence at macroscopic scale, which requires a proper device-independent witness for the validity of our intuitive picture of the everyday macroscopic world. The latter has been formalized in 1985 by Leggett and Garg into a set of assumptions summarized as macrorealism \cite{PhysRevLett.54.857}. 

\begin{mydef}\label{def:macrorealism}
    A macrorealistic system $S$ satisfies the two following assumptions:
    \begin{itemize}
\item
\textbf{Macrorealism per s\'e}: given a set of available macroscopically distinct states, $S$ is in one of them at any given time,
\item
\textbf{Non-invasive measurability}: it is possible in principle to determine the state of $S$ without affecting neither its state nor its dynamical evolution.
\end{itemize}
\end{mydef}
\noindent
Similarly to local realism and Bell inequalities, macrorealism can be consolidated in a set of quantitative conditions which can be experimentally falsified. Given a physical system $S$ evolving in time, one can perform a series of $N$ dichotomic measurements of a macroscopic observable $\mathbb{O}$ at times $t_0, t_1, ..., t_{N-1}$, respectively. Therefore, each measurement produces a random outcome $O_i := O(t_i) \in \{ -1, 1\}$. It can be demonstrated that, if a system $S$ is macrorealistic in accordance with Definition \ref{def:macrorealism}, the statistics of outcomes produced by the considered series of measurements has to satisfy a certain set of inequalities bounding the corresponding correlations and known as Leggett--Garg inequalities. In the simplest case of $N=3$ repeated measurements, the set of LGIs is given as in the following definition~\cite{PhysRep2014,Halliwell2016}.

\medskip
\begin{mydef}\label{def:LGI}
    Given a series of measurements of an observable $\mathbb{O}$ at times $t_0, t_1, t_2$, the corresponding set of \textbf{Leggett--Garg inequalities (LGIs)} is given by
\bea
\label{lgi0121}
  \mathcal{L}_1(t_0, t_1, t_2) = 1+ C_{01} + C_{12} + C_{02} \geq 0 \, , \\[2mm]
\label{lgi0122}
  \mathcal{L}_2(t_0, t_1, t_2) = 1- C_{01} - C_{12} + C_{02} \geq 0 \, , \\[2mm]
  \label{lgi0123}
  \mathcal{L}_3(t_0, t_1, t_2) = 1+ C_{01} - C_{12} - C_{02} \geq 0 \, , \\[2mm]
   \label{lgi0124}
  \mathcal{L}_4(t_0, t_1, t_2) = 1- C_{01} + C_{12} - C_{02} \geq 0 \, ,
\eea
where $C_{ij} := \langle O_i O_j \rangle = \sum_{O_i, O_j \in \{ -1, 1\}} O_i O_j P(O_i, O_j)$ is a correlation function for random outcomes $O_i, O_j$ of measurements at $t_i$ and $t_j$, respectively, and $P(O_i, O_j)$ is the corresponding joint probability distribution.
\end{mydef}
\noindent
Observations of violation of LGIs~(\ref{lgi0121})-\eqref{lgi0124} suggest a failure of the macrorealistic view on the system $S$ in terms of Definition \ref{def:macrorealism}, and can hence be seen as a witness of its ``quantumness" in the sense of ability of outcome statistics predicted by quantum mechanics to violate LGIs.

The LGIs introduced in Definition \ref{def:LGI} are not the unique conditions that can be derived from assumptions of macrorealism. Indeed, similarly to Bell inequalities in the Wigner form, other sets of conditions on the outcome statistics were proposed in literature, which bound directly the joint outcome probabilities $P(O_i, O_j)$ instead of the corresponding correlation functions \cite{Kumari2017,Saha2015}. The idea which lies behind them is that macrorealistic outcome statistics implies the existence of an overall joint probability distribution for outcomes at all times $t_0, ..., t_{N-1}$. Hence, e.g., for $N=3$, the probability distributions $P(O_i, O_j)$ could be computed by marginalizing the overall probability distributions $P(O_0,O_1,O_2)$. From the obvious condition $P(O_0,O_1,O_2) \geq 0$, we develop the following definition of Leggett-Garg inequalities in Wigner form \cite{Naikoo:2019gme}.

\begin{mydef}
    Given a series of measurements of an observable $\mathbb{O}$ at times $t_0, t_1, t_2$, the corresponding set of \textbf{Leggett--Garg inequalities in Wigner form (WLGIs)} is given by
    \begin{widetext}
    \bea \label{wlgi}
        \mathcal{W}_1(t_0, t_1, t_2) & = & P(O_1,O_2)-P(-O_0,O_1)-P(O_0,O_2) \ \leq \ 0 \, , \\[2mm]
        \mathcal{W}_2(t_0, t_1, t_2) & = & P(O_0,O_2)-P(O_0,-O_1)-P(O_1,O_2) \ \leq \ 0 \, , \label{wlgi2} \\[2mm]
        \mathcal{W}_3(t_0, t_1, t_2) & = & P(O_0,O_1)-P(O_1,-O_2)-P(O_0,O_2) \ \leq \ 0 \, , \label{wlgi3}
    \eea
    \end{widetext}
    where $-O_i$ is the measurement outcome alternative to $O_i$, i.e., $-O_i = \mp 1$ if $O_i = \pm 1$.
\end{mydef}
\noindent
Although obvious analogies between macrorealism and local realism (and, in turn, LGIs/WLGIs and CHSH/Bell inequalities) can be drawn, there is a crucial difference between them. For local realism, Bell inequalities provide a necessary and sufficient condition guaranteed by Fine's theorem \cite{Fine1982, halliwellpla}. However, this is not the case for macrorealism: indeed, it has been proven that all forms of LGIs are necessary but not sufficient for macrorealism \footnote{{Henceforth, we refer to macrorealism in its strong form. Indeed, there exists a weaker notion of macrorealism with necessary and sufficient conditions provided by an augmented set of two- and three-time LGIs \cite{Halliwell2017}. Nonetheless, the three-time LGIs alone do not guarantee the validity of a macrorealistic view even in this case.}}, and no counterpart of Fine's theorem exists in this case~\cite{Clemente2015, Clemente2016}. Therefore, while a violation of LGIs/WLGIs guarantees incompatibility of the underlying physical system with the macrorealistic view, their fulfillment can still hide its quantum nature. Nevertheless, a test which completely characterizes macrorealism can be established as two sets of \textit{equalities} putting constraints on the outcome probability distributions, one constraining the signaling from past to future (dubbed no-signaling-in-time) and one constraining signaling from future to past (dubbed arrow-of-time)~\cite{Clemente2015}.
\medskip
\begin{mydef}\label{def:NSIT_AoT}
Given a series of measurements of an observable $\mathbb{O}$ at times $t_0, t_1, t_2$, the corresponding set of \textbf{no-signaling-in-time (NSIT)} conditions is given by:
\bea \label{nsit1}
&& \mathrm{NSIT}^{(1)}: \ \ P(O_2) \ = \ \sum_{O_1} P(O_1, O_2) \, , \\[2mm] \label{nsit2}
&& \mathrm{NSIT}^{(2)}: \ \ P(O_0, O_2) \ = \ \sum_{O_1} P(O_0, O_1, O_2) \, , \\[2mm]
&& \mathrm{NSIT}^{(3)}: \ \ P(O_1, O_2) \ = \ \sum_{O_0} P(O_0, O_1, O_2) \, , \label{nsit3}
\eea
and the set of \textbf{arrow-of-time (AoT)} conditions is given by:
\bea
\label{AoT1} && \mathrm{AoT}^{(1)}: \ \ P(O_0, O_1) \ = \ \sum_{O_2} P(O_0, O_1, O_2) \, , \\[2mm] 
\label{AoT2} && \mathrm{AoT}^{(2)}: \ \ P(O_0) \ = \ \sum_{O_1} P(O_0, O_1) \, , \\[2mm]
\label{AoT3} && \mathrm{AoT}^{(3)}: \ \ P(O_1) \ = \ \sum_{O_2} P(O_1, O_2) \, .
\eea
\end{mydef}
\noindent
Then, NSIT/AoT imply LGIs (WLGIs), while the opposite is not true. This was shown in the physically relevant example of neutrino oscillations in Ref. \cite{Blasone:2023iwf}, where all previous conditions were studied. There, it was shown that NSIT/AoT can witness violations of macrorealism when LGIs (WLGIs) do not. For example, this happens for time intervals much longer than the wave-packets coherence time. This fact will be even clearer in the following, where we are going to analyze the various tests of macrorealism in the framework of meson oscillations. We will see that, with the chosen initial condition, no violation of LGIs or WLGIs is experienced, while NSIT/AoT reveal the quantum nature of the phenomenon.
\subsection{Meson oscillations: an overview}
Before proceeding with the analysis of conditions for macrorealism, we briefly review the phenomenology of neutral meson oscillations. For the sake of simplicity, we focus on neutral K-mesons (kaons), i.e., $K^0-\widebar{K}^0$ oscillations, although the described framework and the following results hold true for other types of neutral mesons as well. Moreover, in our discussion we omit the tiny effects of $\mathcal{CP}$ violation, which does not add any relevant feature to the analysis.

Oscillations of neutral kaons can be effectively approached via the non-relativistic Wigner--Weisskopf approximation (WWA), based on the non-Hermitian Hamiltonian
\begin{equation}\label{eq:nonHermHam}
    \hat{H} = \hat{M} - \frac{i}{2}\hat{\Gamma},
\end{equation}
with the eigenstates $|K_i\rangle$, where $i = S, L$ (dubbed as short-lived and long-lived states, respectively), and the corresponding eigenvalues $m_i - \frac{i}{2}\Gamma_i$, with (distinct) definite masses $m_i$ and decay widths $\Gamma_i$. In the above expression, $\hat{M} = \hat{M}^\dagger$ is the mass operator, which covers the unitary part of the dynamics and $\hat{\Gamma} = \hat{\Gamma}^\dagger$ describes the decay. In turn, the dynamics of a neutral kaon produced at $t_0$ can be obtained by solving the Schr\"odinger equation under the effective WWA Hamiltonian (\ref{eq:nonHermHam}),
\begin{equation}\label{eq:SolutionWWA}
    |\psi(t)\rangle = f_S(t) |K_S\ran + f_L(t) |K_L\ran,
\end{equation}
where $f_i(t) = \lan \psi(t=t_0) | K_i \ran e^{-(im_i + \frac{\Gamma_i}{2})(t-t_0)}$.

The physical (flavor) states $|K^0\rangle$ and $|\widebar{K}^0\rangle$ (kaon and antikaon, respectively) are labeled by the strangeness quantum number. Crucially, for the hadronic decays, both can decay via weak interaction into two or three pions. Generally speaking, the flavor states do not coincide\footnote{In our discussion, for the sake of simplicity we omit the tiny effects of violation of the $\mathcal{CP}$ symmetry in neutral kaon oscillations, which provides an asymmetry in the oscillation probabilities in (\ref{koscfor}) and (\ref{akoscfor}): \begin{eqnarray}\nonumber P_{K^0 \rightarrow \widebar{K}^0}(t) &=& \frac{e^{-\Gamma t}}{2} \frac{|1-\varepsilon|}{|1+\varepsilon|}\Bigl( \cosh \Bigl( \frac{\Delta\Gamma t}{2}\Bigr) - \cos (\Delta m t)\Bigr), \\ \nonumber P_{\widebar{K}^0 \rightarrow K^0}(t) &=& \frac{e^{-\Gamma t}}{2} \frac{|1+\varepsilon|}{|1-\varepsilon|}\Bigl( \cosh \Bigl( \frac{\Delta\Gamma t}{2}\Bigr) - \cos (\Delta m t)\Bigr), \end{eqnarray} where $\epsilon$ is a complex parameter quantifying $\mathcal{CP}$ violation. Nevertheless, we show in the following that the oscillation probabilities enter the conditions for macrorealism only as a combination $P_{K^0 \rightarrow \widebar{K}^0}(t) P_{\widebar{K}^0 \rightarrow K^0}(t)$, which is independent of $\epsilon$.} with the mass eigenstates $|K_i\rangle$, but are rather their superpositions \cite{GellMann1955}, that is
\bea
|K^0\ran & = & \frac{1}{\sqrt{2}} \, \lf(|K_S\ran +|K_L\ran\ri) \, , \\[2mm]
|\widebar{K}^0\ran & = & \frac{1}{\sqrt{2}} \, \lf(|K_S\ran -|K_L\ran\ri) \, .
\eea
Now, suppose that a neutral kaon is produced at $t=t_0$ as $|\psi(t=t_0)\ran = |K^0\ran$. Taking into account the dynamics given by (\ref{eq:SolutionWWA}), it evolves into the state
\begin{eqnarray}
\nonumber |K^0(t)\ran &=&  \frac{1}{\sqrt{2}} \Bigl(e^{-(i m_S + \frac{\Ga _S}{2}) \Delta t} |K_S\ran \\
&+& e^{-i (m_L + \frac{\Ga_L}{2}) \Delta t} |K_L\ran \Bigr), \label{eq:EvolvedState}
\end{eqnarray}
where $\Delta t = t - t_0$. Hence, the probabilities of finding a kaon (survival probability) and antikaon (oscillation probability) at time $t$ are given by
\be\label{eq:KaonProbs}
P_{K^0 \rightarrow K^0/\widebar{K}^0}(t) \ = \ |\lan K^0/\widebar{K}^0|K^0(t)\ran|^2 \, ,
\ee
respectively, and it can be straightforwardly calculated from (\ref{eq:EvolvedState}) as
\begin{equation} \label{koscfor}
P_{K^0 \rightarrow K^0 / \widebar{K}^0}(t) = \frac{e^{-\Gamma \Delta t}}{2} \Bigl( \cosh \Bigl( \frac{\Delta\Gamma \Delta t}{2}\Bigr) \pm \cos (\Delta m \Delta t)\Bigr) \, ,
\end{equation}
where $\Delta m = m_L - m_S$ is the difference of neutral kaon masses, $\Gamma = \frac{\Gamma_S + \Gamma_L}{2}$ and $\Delta\Gamma = \Gamma_S - \Gamma_L$. Similarly, the survival and oscillation probabilities can be calculated for the scenario where an antikaon is produced at $t=t_0$, so that $|\psi(t=t_0)\ran = |\widebar{K}^0\ran$ and
\begin{equation} \label{akoscfor}
P_{\widebar{K}^0 \rightarrow \widebar{K}^0 / K^0}(t) = \frac{e^{-\Gamma \Delta t}}{2} \Bigl( \cosh \Bigl( \frac{\Delta\Gamma \Delta t}{2}\Bigr) \pm \cos (\Delta m \Delta t)\Bigr) \, .
\end{equation}
Note that the survival and oscillation probabilities do not sum up to unity, $P_{K^0 \rightarrow \widebar{K}^0}(t)+P_{K^0 \rightarrow K^0}(t) \neq 1$, because of the hadronic and the (semi-)leptonic decays of neutral kaons. This fact strongly distinguishes the present case from the one of neutrino oscillations. In particular, the (semi-)leptonic decays were intensively used to study neutral kaon oscillations at accelerator facilities in the CPLEAR experiment. Therein, the flavor of the decayed particle is uniquely identified by a lepton of a definite charge, and the relative decay rates provide a direct measure of the relative flavor components of the particle's state. In turn, the transition probabilities can be associated with the corresponding decay rates.
\section{Macrorealism in meson oscillations} \label{sec:3}

\subsection{Oscillation probabilities}

In order to characterize the quantumness of neutral kaon oscillations as done in Section \ref{sec:2A}, it is necessary to define an observable that can be associated with the flavor of the particle. This can be done by choosing a dichotomic observable $\mathbb{O}^{F}$ represented by the corresponding operator
\begin{equation}\label{eq:obs}
    \hat{O}^{F} = 2 \Pi_{F} - \mathds{1},
\end{equation}
where $\Pi_{F} = |F\rangle\langle F|$ is a projector on the state of flavor $F \in \{K^0, \widebar{K}^0\}$. A measurement of $\mathbb{O}^{F}$ at time $t_i$ thus reveals whether the neutral meson is found at $t_i$ in flavor $F$ or not\footnote{{This choice of the dichotomic observable $\mathbb{O}$ is motivated by the incomplete description of a neutral kaon system when considering the Hilbert space spanned by the flavor states $|K^0\rangle$ and $|\widebar{K}^0\rangle$. Indeed, this space does not include states associated with the decay products.}}. Therefore, we associate its possible outcomes $\{+1, -1\} \ni O_i^{F}$ with symbols $F$ and $\neg F$. This allows us to introduce probabilities $P(F|t_i) \equiv P(O_i = +1)$ and $P(\neg F|t_i) \equiv P(O_i = -1)$ of events "particle is found in flavor $F$" and "particle is not found in flavor $F$", respectively. A crucial difference between this scenario and two-flavor neutrino oscillations can be easily spotted: for the latter, a negative outcome of measurement of $\mathbb{O}^{F}$ means that the neutrino is found in the conjugated flavor $\widebar{F}$ (e.g., if $F = \nu_e$ is an electronic neutrino, then $\widebar{F} = \nu_\mu$ is a muonic netrino). This is not the case for neutral kaon oscillations, because the outcome $\neg F$ includes decay events as well.

Before moving on with the derivation of conditions for macrorealism for neutral kaon oscillations, we derive the necessary joint probabilities for outcomes of sequential measurements. As neutral mesons are produced in accelerator facilities in a state of given flavor, we assume the initial state to be one which, without loss of generality, is chosen as $|F\rangle$. Moreover, as we are interested in quantification in terms of experimentally accessible quantities, we aim in connecting the joint outcome probabilities with transition probabilities (\ref{eq:KaonProbs}). This is achieved in the following Lemma.

\begin{Lemma}\label{lem:JointProbs}
    For a neutral kaon produced at $t_0$ in the flavor $F$, two measurements of the observable $\mathbb{O}^{F}$ performed at times $t_1$ and $t_2$, respectively, reveal outcomes $O_1 \in \{ F, \neg F\}$ and $O_2 \in \{ F, \neg F\}$ with probabilities
\begin{eqnarray}
    \nonumber P(F, F) &=& P_{F \rightarrow F}(\Delta t_1) P_{F \rightarrow F}(\Delta t_2), \\
    \nonumber P(F, \neg F) 
    &=& P_{F \rightarrow F}(\Delta t_1) \Bigl( 1-P_{F \rightarrow F}(\Delta t_2)\Bigr), \\
    \nonumber P(\neg F, F) 
    &=& P_{F \rightarrow \widebar{F}}(\Delta t_1) P_{\widebar{F} \rightarrow F}(\Delta t_2 ), \\
    \nonumber P(\neg F, \neg F) 
    &=& 1 - P_{F \rightarrow F}(\Delta t_1) - P_{F \rightarrow \widebar{F}}(\Delta t_2) P_{\widebar{F} \rightarrow F}(\Delta t_2),
\end{eqnarray}
where $\Delta t_1 = t_1 - t_0$ and $\Delta t_2 = t_2 - t_1$, and $\widebar{F}$ is the flavor conjugated to $F$, so that $\widebar{\widebar{K}^0} = K^0$, and $P(\cdot, \cdot) := P(\cdot, \cdot | t_1, t_2)$.
\begin{proof}
    See Appendix \ref{app:A}.
\end{proof}
\end{Lemma}

\subsection{(W)LGIs for neutral kaon system}
\begin{figure}[t!]
\hspace{-2.5mm}    \includegraphics[width=0.5\textwidth]{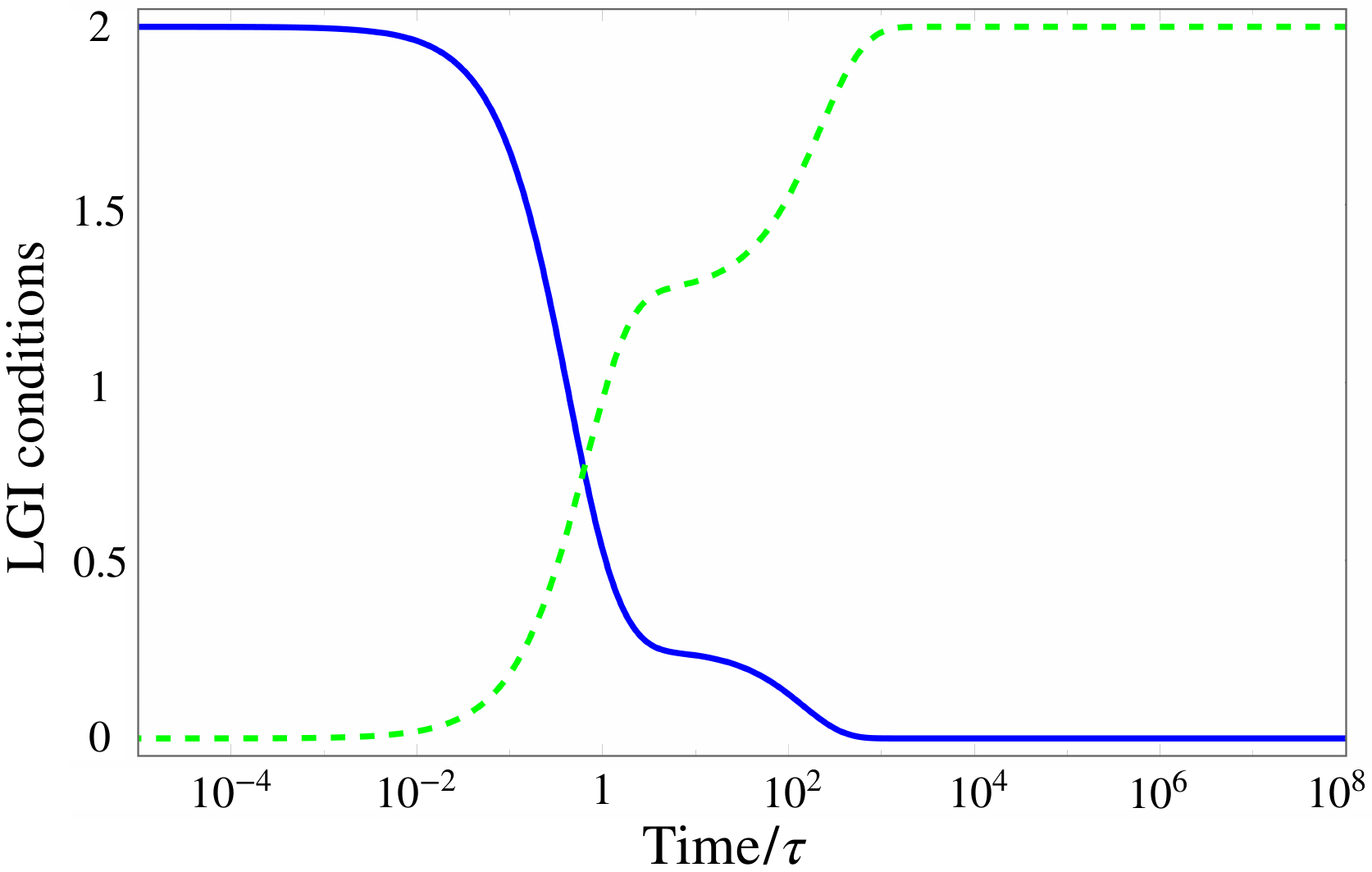}
\hspace{-2.5mm}  \includegraphics[width=0.5\textwidth]{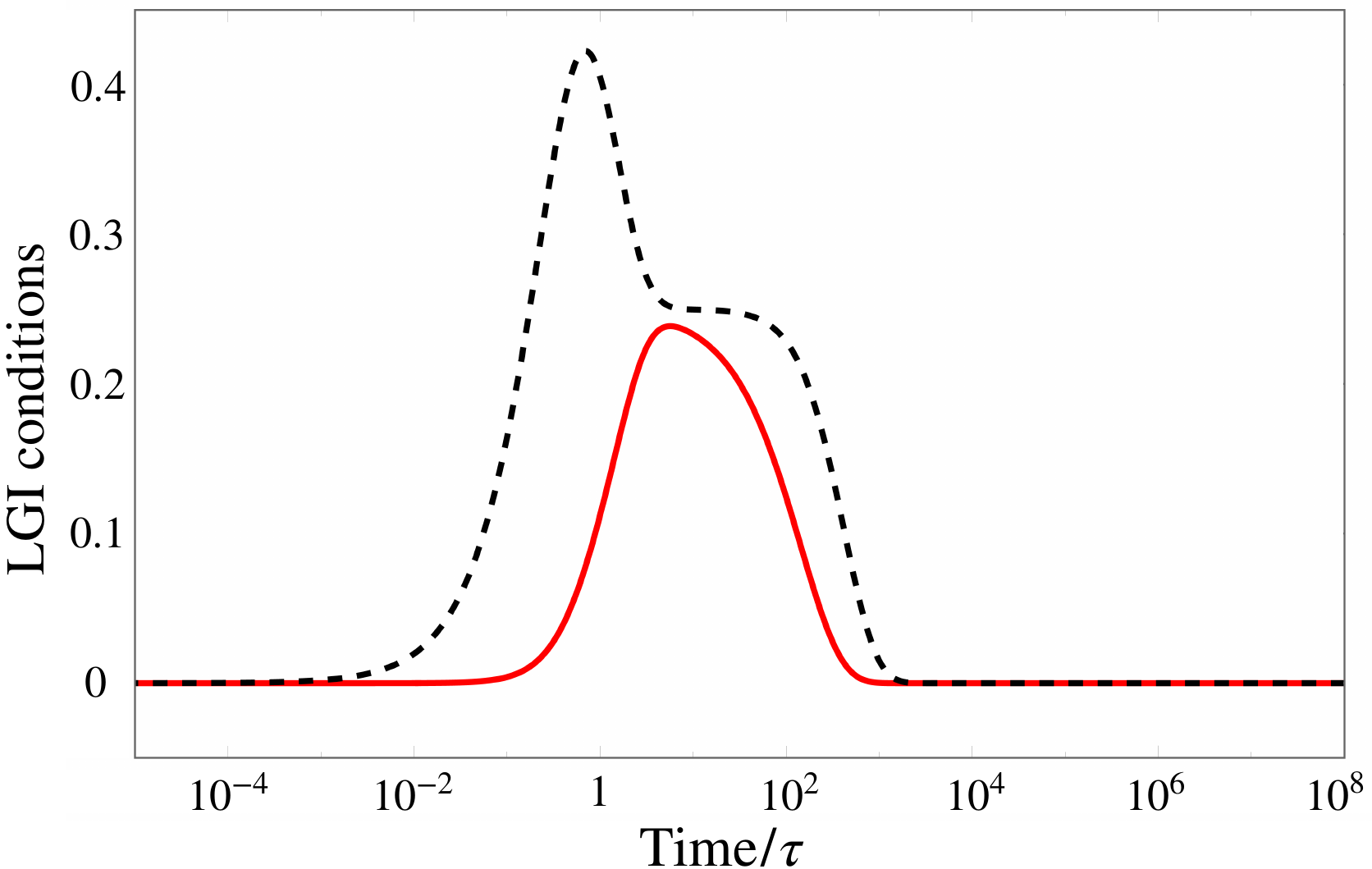}
    \caption{Functions $\mathcal{L}_{1}(t)$ (blue solid curve, first panel), $\mathcal{L}_{2}(t)$ (red solid curve, second panel), $\mathcal{L}_{3}(t)$ (black dashed curve, second panel), and $\mathcal{L}_{4}(t)$ (green dashed curve, first panel) as functions of time scaled by the proper mean lifetime $\tau = 8.954 \cdot 10^{-11} \; \mathrm{s}$ of a neutral kaon. We assume that the kaon is produced in flavor $F = K^0$, and the parameters $\Gamma = 5.5939 \times 10^{9} \; \mathrm{s}^{-1}$, $\Delta \Gamma = 1.1149 \times 10^{10} \; \mathrm{s}^{-1}$, and $\Delta m = 0.5293 \times 10^{10} \hbar \; \mathrm{s}^{-1}$ for neutral kaon system are chosen in accordance with the corresponding experimental values provided by the Particle Data Group in \cite{PDG}. {All the quantities that appear in the plots are dimensionless.}}
\label{lgi1png}
\end{figure}

We start by deriving the Leggett-Garg inequalities in their canonical~(\ref{lgi0121})-\eqref{lgi0124} as well as Wigner form~(\ref{wlgi})-\eqref{wlgi3} for neutral kaon oscillations. In order to address typical experimental setups for neutral meson systems, we assume that the kaon is produced at $t_0=0$, and the measurements of $\mathbb{O}^{F}$ are performed at equidistant time intervals $\Delta t_1 = \Delta t_2 = t$, choosing therefore $t_0 = 0$, $t_1 = t$, $t_2 = 2t$. Having fixed that, we aim at the evaluation of the correlation functions 
\be
C_{ij} \ = \ \sum_{O_i,O_j\in \{F, \neg F\}} \, O_i \, O_j \, P(O_i,O_j) \, ,
\ee
which can be straightforwardly obtained by applying Lemma \ref{lem:JointProbs}, so as to obtain
\begin{eqnarray}
    C_{01} &=& 2P_{F \rightarrow F}(t) - 1, \\
    \nonumber C_{12} &=& 1 - 2P_{F \rightarrow F}(t) \Bigl( 1 - P_{F \rightarrow F}(t) \Bigr) \\
    &-& 2  P_{F \rightarrow \widebar{F}}(t) P_{\widebar{F} \rightarrow F}(t), \\
    C_{02} &=& 2P_{F \rightarrow F}(2t) - 1.
\end{eqnarray}
{Substituting} these formulas into~(\ref{lgi0121})-\eqref{lgi0124}, we obtain the set of LGIs constraining neutral kaon transition probabilities:
\begin{eqnarray}\label{eq:LGIKaons1}
  \nonumber \mathcal{L}_1(t) & = &   P_{F \rightarrow F}(2t) + P_{F \rightarrow F}^2(t) \\
  &-&  P_{F \rightarrow \widebar{F}}(t) P_{\widebar{F} \rightarrow F}(t) \ \geq \ 0, \\[2mm]
   \nonumber \mathcal{L}_2(t) & = &    P_{F \rightarrow F}(2t) - P_{F \rightarrow F}^2(t) \\
   &+&  P_{F \rightarrow \widebar{F}}(t) P_{\widebar{F} \rightarrow F}(t)  \ \geq \ 0, \\[2mm]
  \mathcal{L}_3(t)  & = &  -P_{F \rightarrow F}(2t) - P_{F \rightarrow F}^2(t) + 2P_{F \rightarrow F}(t) \non \\[2mm]
	& + & P_{F \rightarrow \widebar{F}}(t) P_{\widebar{F} \rightarrow F}(t) \ \geq \ 0, \\[2mm]
  \label{eq:LGIKaons4} \mathcal{L}_4(t)  & = & - P_{F \rightarrow F}(2t) + P_{F \rightarrow F}^2(t) + 2 \Bigl(1 - P_{F \rightarrow F}(t)\Bigr) \non \\[2mm]
	& - &  P_{F \rightarrow \widebar{F}}(t) P_{\widebar{F} \rightarrow F}(t) \ \geq \  0.
\end{eqnarray}
Without loss of generality, assuming $F = K^0$ and applying the transition probabilities (\ref{eq:KaonProbs}), we plot the functions~(\ref{eq:LGIKaons1})-(\ref{eq:LGIKaons4}) in Fig.~\ref{lgi1png}. In contrast to neutrino oscillations \cite{Blasone:2023iwf}, the entire set of LGIs provide four different constraints for neutral kaon oscillations. Nevertheless, it can be easily noted that, at any point in time, they are not violated. Therefore, LGIs cannot detect violation of macrorealism in neutral kaon system and could lead to an erroneous idea that its oscillations are susceptible of a macrorealistic interpretation despite their quantum nature.

{Before continuing, it must be pointed out that violations of LGIs are closely related to the choice of observables and initial conditions. For instance, in Ref. \cite{fmeson} it is possible to find that, by selecting the strangeness as the dichotomic observable and considering four-time LGIs (instead of the three-time LGIs), one comes across a violation of the inequalities even when dealing with meson oscillations. This result must not be surprising, as there are no ambiguities, overlaps, and confusions between the framework discussed in Ref. \cite{fmeson} and the one we are considering here; indeed, they are covering essentially different scenarios which are not compatible, and thus not comparable. Our choice to work with three-time LGIs is simply dictated by the fact that our purpose is to exhibit the incompleteness of the LGIs macrorealistic description with respect to the necessary and sufficient NSIT conditions, which so far have been explored efficiently for three distinct measurement times only.}

\begin{figure}[t!]
   \centering
\hspace{-7mm}    \includegraphics[width=0.5\textwidth]{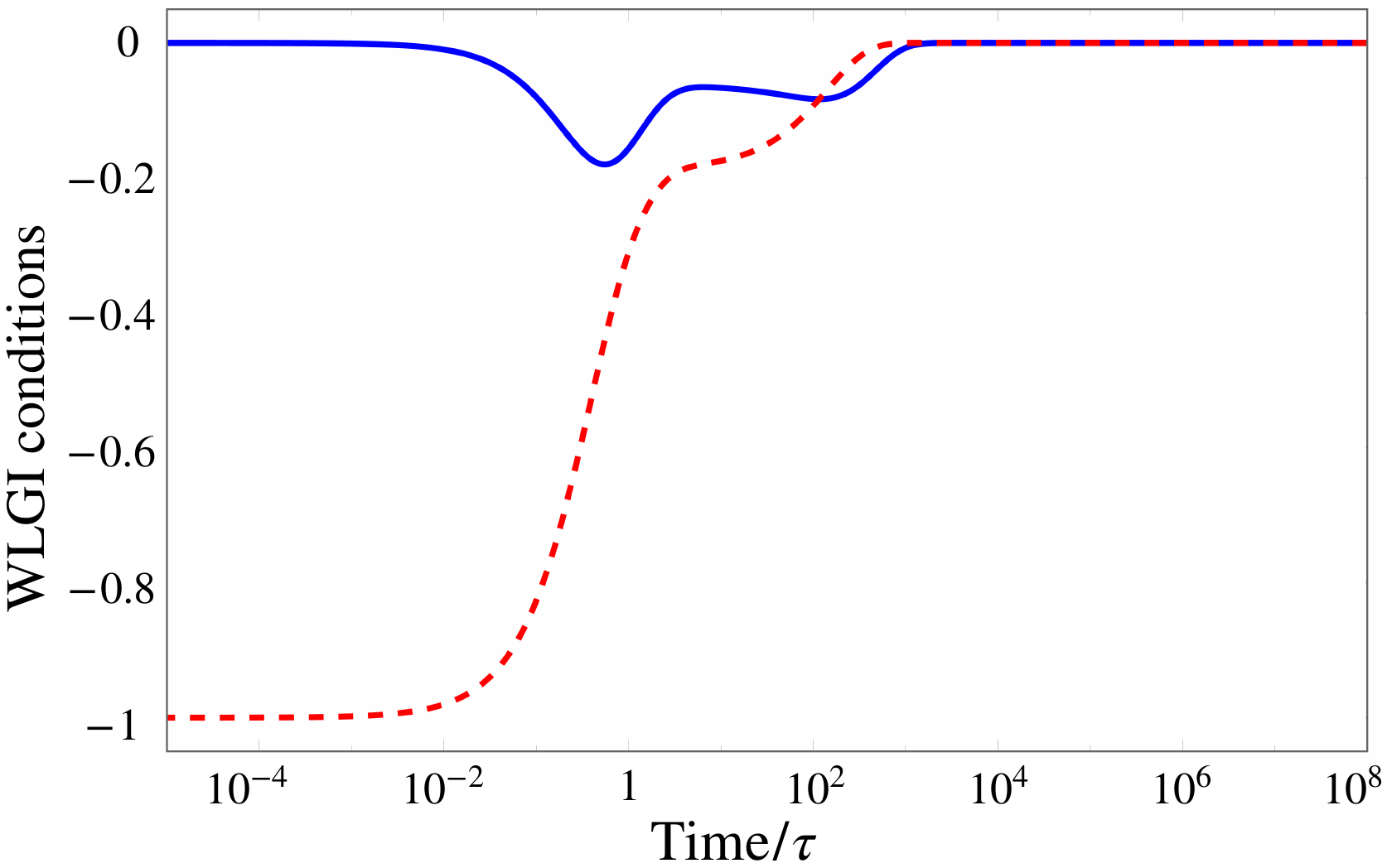}
    \caption{Functions $\mathcal{W}_{1}(t)$ (blue solid curve) and $\mathcal{W}_{2}(t)$ (red dashed curve) as functions of time scaled by the proper mean lifetime $\tau = 8.954 \cdot 10^{-11} \; \mathrm{s}$ of a neutral kaon. We assume that the kaon is produced in flavor $F = K^0$, and the parameters $\Gamma = 5.5939 \times 10^{9} \; \mathrm{s}^{-1}$, $\Delta \Gamma = 1.1149 \times 10^{10} \; \mathrm{s}^{-1}$, and $\Delta m = 0.5293 \times 10^{10} \hbar \; \mathrm{s}^{-1}$ for neutral kaon system are chosen in accordance with the corresponding experimental values provided by the Particle Data Group in \cite{PDG}. {All the quantities that appear in the plot are dimensionless.}}
    \label{wlgispng}
\end{figure}

Next, we proceed with questioning the suitability of WLGIs for the characterization of the quantumness of neutral kaon oscillations. WLGIs (\ref{wlgi})-\eqref{wlgi3} directly depend on probabilities of measurement outcomes which are fixed. Therefore, we assume $O_0 = F$, $O_1 = \neg F$, and $O_2 = \neg F$, so that $-O_0 = \neg F$, $-O_1 = F$, and $-O_2 = F$. Applying Lemma \ref{lem:JointProbs}, we obtain the following set of WLGIs constraining neutral kaon transition probabilities:
\begin{eqnarray}
   \nonumber \mathcal{W}_1(t) & = & P_{F \rightarrow F}(2t) - P_{F \rightarrow F}(t) \\
   &-&  P_{F \rightarrow \widebar{F}}(t) P_{\widebar{F} \rightarrow F}(t) \leq 0, \\[2mm]
    \mathcal{W}_2(t) & = &  P_{F \rightarrow \widebar{F}}(t) P_{\widebar{F} \rightarrow F}(t) - P_{F \rightarrow F}(2t) \leq 0, \\
   \nonumber \mathcal{W}_3(t) & = & P_{F \rightarrow F}(2t) - P_{F \rightarrow F}(t) \\
   &-&  P_{F \rightarrow \widebar{F}}(t) P_{\widebar{F} \rightarrow F}(t) \leq 0.
\end{eqnarray}
Therefore, WLGIs provide two different conditions requiring negativity of $\mathcal{W}_1(t)$ and $\mathcal{W}_2(t)$, which are plotted in Fig. \ref{wlgispng} under the assumption $F = K^0$. Similarly to the LGIs analyzed above, we find that WLGIs are satisfied at every point of time, demonstrating a crucial difference with WLGIs for neutrino oscillations, which reveal violation of macrorealism at small times \cite{Blasone:2023iwf}. This means that (W)LGIs are not efficient enough to test the quantum nature of flavor transitions in neutral kaon oscillations, and hence the NSIT/AoT (as necessary and sufficient conditions for macrorealism) have to be addressed. 

\subsection{Necessary and sufficient conditions for macrorealism in neutral kaon oscillations}

\begin{figure}[t!]
   \centering
\hspace{-4mm}    \includegraphics[width=0.5\textwidth]{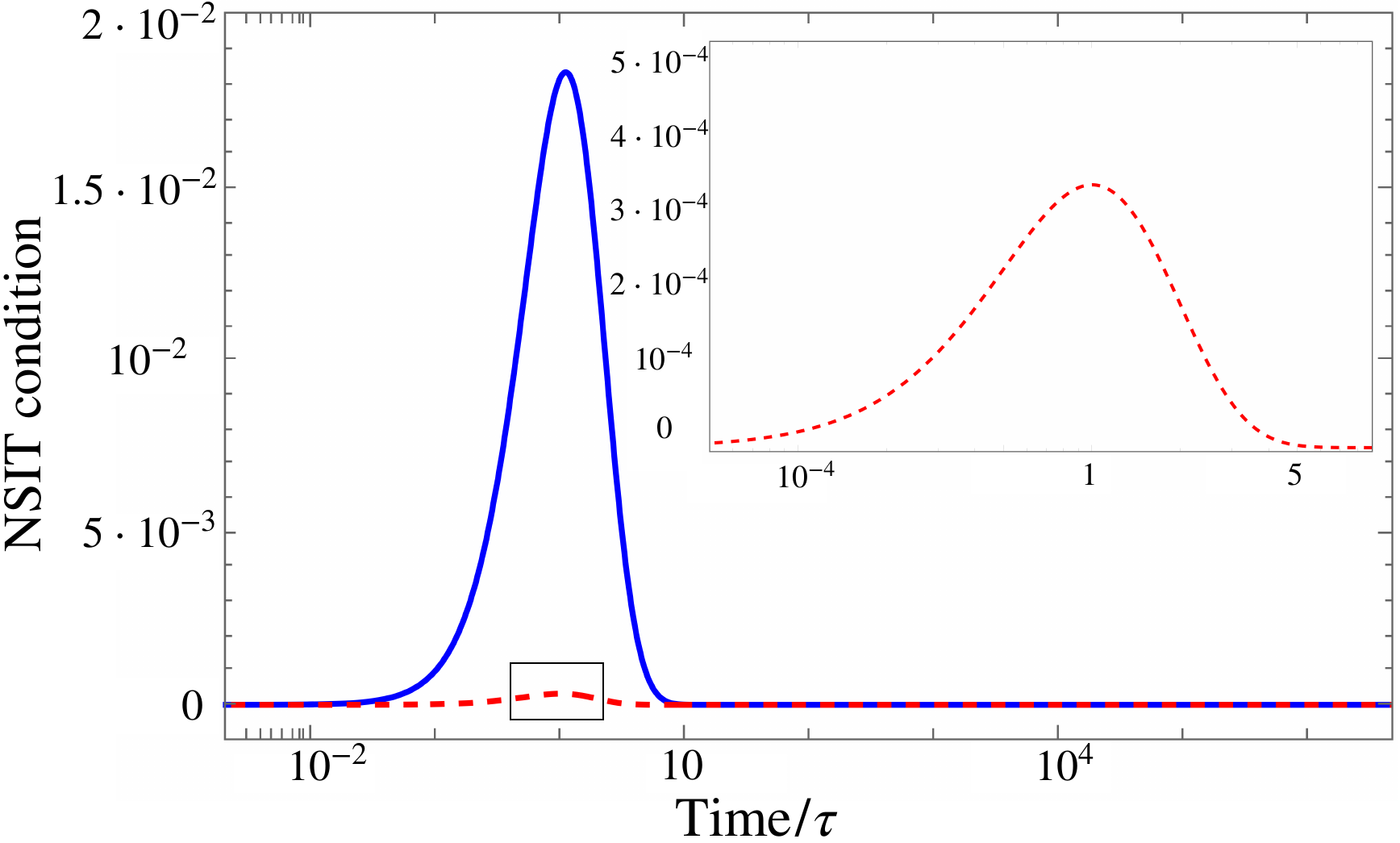}
    \caption{Function $\mathcal{N}(t)$ of the neutral kaon (blue solid curve)  and of the strange B meson (red dashed curve) as functions of time scaled by the proper mean lifetime $\tau = 8.954 \cdot 10^{-11} \; \mathrm{s}$ for a neutral kaon and $\tau = 1.470 \cdot 10^{-12} \; \mathrm{s}$ for a strange B meson. We assume that the kaon is produced in flavor $F = K^0$, and the parameters $\Gamma = 5.5939 \times 10^{9} \; \mathrm{s}^{-1}$, $\Delta \Gamma = 1.1149 \times 10^{10} \; \mathrm{s}^{-1}$, and $\Delta m = 0.5293 \times 10^{10} \hbar \; \mathrm{s}^{-1}$ for neutral kaon system are chosen in accordance with the corresponding experimental values provided by the Particle Data Group in \cite{PDG}. For the strange B meson, the particle is produced in the flavor $F=B_s$ and the parameters are $\Gamma = 6.615 \times 10^{11} \; \mathrm{s}^{-1}$, $\Delta \Gamma = 9.14 \times 10^{10} \; \mathrm{s}^{-1}$, and $\Delta m = 1.776 \times 10^{13} \hbar \; \mathrm{s}^{-1}$. {All the quantities that appear in the plot are dimensionless.}}
    \label{fig:my_nsitvslgi2}
\end{figure}

Similarly to WLGIs, the NSIT/AoT conditions depend directly on the probabilities of measurement outcomes. Nevertheless, only some of the latter are fixed: therefore, we assume $O_0 = F$, $O_1 = \neg F$, and $O_2 = \neg F$ as above, unless there is summation over the corresponding measurement outcome. Applying Lemma \ref{lem:JointProbs}, we find that the set of AoT conditions is trivially satisfied, while the set of NSIT conditions reduces to a unique, non-trivial condition with respect to \eqref{nsit1}, which is equal to
\begin{equation}\label{eq:RawKaonNSIT}
    1 - P_{F \rightarrow F}(2t) = 1 - P_{F \rightarrow F}^2(t) -  P_{F \rightarrow \widebar{F}}(t) P_{\widebar{F} \rightarrow F}(t).
\end{equation}
For the sake of simplicity, we use (\ref{eq:RawKaonNSIT}) to introduce the function
\begin{equation} \label{ntnsitdef}
\mathcal{N}(t) :=  P_{F \rightarrow F}(2t) - P_{F \rightarrow F}^2(t) - P_{F \rightarrow \widebar{F}}(t) P_{\widebar{F} \rightarrow F}(t),
\end{equation}
so that the NSIT condition for macrorealism in neutral kaon oscillations has the simple form
\be \label{ntnsit}
\mathcal{N}(t) \ = \ 0 \, .
\ee
Without loss of generality, assuming $F = K^0$ and applying the transition probabilities (\ref{eq:KaonProbs}), we plot the function~(\ref{ntnsitdef}) in Fig.~\ref{fig:my_nsitvslgi2}. The plot shows a strong contrast between (W)LGIs and NSIT/AoT conditions in the case of neutral kaon oscillations: while the former are never violated (thus failing to catch the quantum nature of flavor oscillations), the latter are violated at any time point excluding the trivial cases $t=0$ and $t \rightarrow \infty$. This suggests the complete incompatibility of neutral kaon systems with the macrorealistic view.

Nevertheless, it is necessary to remark that the lack of violation of LGIs and WLGIs seems to be contingent on the specific choice of initial conditions, i.e. the production of neutral kaon in flavor $K^0$ at $t=0$. Other choices could lead to violation of (W)LGIs \cite{Naikoo2018B}. However, the strong difference between  predictions of (W)LGIs and NSIT/AoT conditions discovered so far clearly highlights the limitations of the former as tools for capturing the quantum nature of the considered phenomenon.

{Let us conclude with a final remark about the possibility of experimentally testing violations of NSIT/AoT. In Refs. \cite{Formaggio2016,Fu:2017hky}, the data from the MINOS and Daya Bay experiments were analyzed to test LGIs. Starting from the experimental best fits, a set of pseudo-data was generated with a Gaussian distribution. The violations of LGIs were then compared to those arising from statistical fluctuations (false positives) in a macrorealistic model, leading to an observed statistical significance of $>5\sigma $. In principle, a similar approach could be repeated in the present case for NSIT/AoT.}
\section{Discussion and Outlook}
\label{sec:4}
In this study, we have undertaken an analysis of NSIT/AoT conditions within the physical framework of meson oscillations. Our study reveals that, just like the scenario occurred for two-flavor neutrino oscillations, these conditions boil down to a singular, non-trivial equality, which is commonly violated by meson oscillations. Employing the specific choice made for initial conditions, it is observed that the LGIs and WLGIs are always fulfilled, while NSIT/AoT reveal a violation of macrorealism. These findings unambiguously underscore the efficacy of such a formalism in scrutinizing the intricate interplay between particle physics and the foundational aspects of quantum mechanics.

A significant step forward along this direction would involve the investigation of NSIT/AoT conditions in quantum field theory (QFT). In Ref. \cite{Blasone:2023mbc}, WLGIs were investigated within the framework of the \emph{flavor Fock-space} approach of QFT neutrino physics, allowing for a consistent description of flavor oscillations at all energy scales. The results of that study indicate that QFT appears even more incompatible with macrorealism than quantum mechanics. This finding is in agreement with similar results on the radically nonclassical structure of QFT. For instance, in the context of quantum nonlocality, the vacuum state in QFT generically features maximum violation of the Bell inequalities \cite{Summers:1987fepr,Summers:1987fn,Summers:1987ze}; moreover, Bell inequalities can be violated beyond the standard Tsirelson in the quantum mechanical limit of phenomenological models of quantum graivty incorporating a minimal length \cite{Bosso2023}. Based on these considerations, given that NSIT/AoT conditions are necessary and sufficient for macrorealism (akin to Bell inequalities for local realism), we believe they could contribute to unveiling foundational characteristics of quantum field theory and, further on, models of quantum gravity with or without a fundamental scale of length. On a final note, we also pinpoint that the flavor vacuum and its quantum information properties were studied in the case of boson mixing as well \cite{Blasone:2001du,bigs2,ill6}, so that the analysis performed in this paper can also be extended to the case of QFT.

\section*{Acknowledgements}
F. I. acknowledges support from the European Union – Next Generation EU through Project PRIN 2022-PNRR no. P202253RLY “Harnessing topological phases for quantum technologies”. L. P. acknowledges networking support by the COST Action CA18108. K.~S. acknowledges that this research was funded in whole or in part by the Austrian Science Fund (FWF) 10.55776/PAT4559623.


\appendix

\section{Proof of Lemma \ref{lem:JointProbs}}\label{app:A}

Given a quantum system evolving in time under the initial condition $\rho(t=t_0) = \rho_0$ for its state, a joint probability of getting outcomes $O_1$ and $O_2$ of measurements of an observable $\mathbb{O}$ at $t_1$ and $t_2$, respectively, is given by:
\begin{equation}
    P(O_1, O_2 | t_1, t_2) = \operatorname{Tr}\Biggl[ \Pi_{O_2} V_{t_2 - t_1} \Bigl[ \Pi_{O_1} V_{t_1 - t_0}[ \rho_0 ] \Pi_{O_1} \Bigr] \Biggr],
\end{equation}
where $\Pi_{O_i}$ is a projector onto an eigenspace associated with the measurement outcome $O_i$, and $V_{t_1 - t_0}[\rho(t_0)] = \rho(t_1)$ is a dynamical map describing the time evolution of the system. 

For the sake of consistency, we recall that the dynamics of a decaying particle system, such as a neutral kaon system described via the effective WWA non-Hermitian Hamiltonian (\ref{eq:nonHermHam}), can be equivalently described as an open system dynamics via the following Gorini-Kossakowski-Lindblad-Sudarshan (GKLS) evolution equation for a state $\varrho$ on an extended Hilbert space $\mathbf{H} = \mathbf{H}_f \oplus \mathbf{H}_d$, where $\mathbf{H}_f$ is the flavor Hilbert space spanned by $\{ |K^0\rangle, |\widebar{K}^0\rangle\}$, and $\mathbf{H}_d$ is a Hilbert space spanned by states $\{|d_k\rangle\}_k$ corresponding to decay products \cite{Bertlmann2006}:
\begin{equation}\label{eq:app:TotalDyn}
    \dot{\varrho} = -i[\mathcal{M}, \varrho ] - \frac{1}{2}\Bigl(\mathcal{B}^\dagger \mathcal{B} \varrho + \varrho \mathcal{B}^\dagger \mathcal{B} - 2 \mathcal{B} \varrho \mathcal{B}^\dagger\Bigr).
\end{equation}

Due to the tensor sum structure of $\mathbf{H}$, the total state can be decomposed as $\varrho = \begin{pmatrix} \rho_f & \rho_{fd} \\ \rho_{fd}^\dagger & \rho_d \end{pmatrix}$, and we define
\begin{equation}
    \mathcal{M} = \begin{pmatrix} M & 0 \\ 0 & 0\end{pmatrix}, \;
    \mathcal{B} = \begin{pmatrix} 0 & 0 \\ B & 0\end{pmatrix},
\end{equation}
where $M$ is the mass operator of the WWA Hamiltonian (\ref{eq:nonHermHam}), while $B = \sum_{kj} b_{kj} |d_k\rangle\langle f_j|$, with $\{|f_j\rangle\}_j$ spanning the flavor Hilbert space $\mathbf{H}_f$, is an operator mapping states from $\mathbf{H}_f$ onto $\mathbf{H}_d$, thus covering the decay property \cite{Bertlmann2006}. The GKLS evolution equation (\ref{eq:app:TotalDyn}) can be decomposed into three dynamical equations:
\begin{eqnarray}\label{app:eq:rhof}
    \dot{\rho}_f &=& -i[M, \rho_f ] - \frac{1}{2}\{ \Gamma, \rho_f \}, \\
    \dot{\rho}_{fd} &=& -i M \rho_{fd} - \frac{1}{2}\Gamma \rho_{fd}, \\
    \dot{\rho}_d &=& B \rho_f B^\dagger . \label{app:eq:rhod}
\end{eqnarray}

In the above, $\Gamma = B^\dagger B = \sum_{kjj'} b_{kj}^* b_{kj'} |f_j\rangle\langle f_{j'}|$. It can be associated with the non-Hermitian term $\Gamma$ of the WWA Hamiltonian (\ref{eq:nonHermHam}) by an appropriate choice of $b_{kj}$ and $\{|f_j\rangle\}_j$, so that (\ref{app:eq:rhof}) coincides with the Schr\"odinger equation under the WWA Hamiltonian (\ref{eq:nonHermHam}).

As we assume that a neutral kaon is produced at $t_0$ in a certain flavor state $|F\rangle \in \mathbf{H}_f$, it is easy to spot that the component $\rho_{fd}$ remains zero at every point in time, so that $\varrho(t) = \rho_f(t) \oplus \rho_d(t)$. Now, associating the dynamical map $V_t[\cdot]$ with dynamical equations (\ref{app:eq:rhof}) and (\ref{app:eq:rhod}), and outcomes $F$ and $\neg F$ of measurements of the observable (\ref{eq:obs}) with projectors $\Pi_F = |F\rangle \langle F|$ and $\mathds{1} - \Pi_F$, respectively, we calculate first the joint probabilities of finding $F$ in the first measurement and $F$/$\neg F$ in the second measurement, which are straightforward:
\begin{eqnarray}
    \nonumber P(F,F) &=& \operatorname{Tr}\Biggl[\Pi_F V_{t_2 - t_1}\Bigl[ \Pi_F V_{t_1 - t_0}[\Pi_F ] \Pi_F \Bigr]\Biggr] \\
    \nonumber &=& \langle F | V_{t_1 - t_0}[\Pi_F ] | F \rangle \operatorname{Tr}\Bigl[\Pi_F V_{t_2 - t_1}[ \Pi_F ]\Bigr] \\
    \nonumber &=& \langle F | V_{t_1 - t_0}[\Pi_F ] | F \rangle \langle F | V_{t_2 - t_1}[\Pi_F ] | F \rangle \\
    &=& P_{F \rightarrow F}(t_1 - t_0) P_{F \rightarrow F}(t_2 - t_1), \\
    \nonumber P(F,\neg F) &=& \operatorname{Tr}\Bigl[(\mathds{1} - \Pi_F) V_{t_2 - t_1}[ \Pi_F V_{t_1 - t_0}[\Pi_F ] \Pi_F ]\Bigr] \\
    \nonumber &=& \langle F | V_{t_1 - t_0}[\Pi_F ] | F \rangle \operatorname{Tr}\Bigl[(\mathds{1} - \Pi_F) V_{t_2 - t_1}[ \Pi_F ]\Bigr] \\
    \nonumber &=& \langle F | V_{t_1 - t_0}[\Pi_F ] | F \rangle \Bigl( 1 - \langle F | V_{t_2 - t_1}[\Pi_F ] | F \rangle\Bigr) \\
    &=& P_{F \rightarrow F}(t_1 - t_0) \Bigl( 1 - P_{F \rightarrow F}(t_2 - t_1)\Bigr).
\end{eqnarray}

\noindent
On the other hand, for the probability $P(\neg F,F)$, we take into account that the projector onto the eigenspace of $\neg F$ can be given as
\begin{equation}
    \mathds{1} - \Pi_F = \Pi_{\widebar{F}} + \Pi_d = | \widebar{F} \rangle \langle \widebar{F} | + \Pi_d,
\end{equation}
where $\Pi_d$ is a projector onto $\mathbf{H}_d$. Hence, we obtain
\begin{eqnarray}
    \nonumber P(\neg F,F) &=& \operatorname{Tr}\Biggl[\Pi_F V_{t_2 - t_1}\Bigl[ (\mathds{1} - \Pi_F) V_{t_1 - t_0}[\Pi_F ] (\mathds{1} - \Pi_F) \Bigr]\Biggr] \\
    \nonumber &=& \operatorname{Tr}\Biggl[\Pi_F V_{t_2 - t_1}\Bigl[ (\Pi_{\widebar{F}} + \Pi_d) V_{t_1 - t_0}[\Pi_F ] (\Pi_{\widebar{F}} + \Pi_d) \Bigr]\Biggr] \\
    \nonumber &=& \operatorname{Tr}\Biggl[\Pi_F V_{t_2 - t_1}[ \Pi_{\widebar{F}} V_{t_1 - t_0}[\Pi_F ] \Pi_{\widebar{F}} + \Pi_d V_{t_1 - t_0}[\Pi_F ] \Pi_d ]\Biggr] \\
    \nonumber &=& \operatorname{Tr}\Bigl[\Pi_F V_{t_2 - t_1}[ \Pi_{\widebar{F}} V_{t_1 - t_0}[\Pi_F ] \Pi_{\widebar{F}} ]\Bigr] \\
    \nonumber &=& \langle \widebar{F} | V_{t_1 - t_0}[\Pi_F ] | \widebar{F} \rangle \operatorname{Tr}\Bigl[\Pi_F V_{t_2 - t_1}[ \Pi_{\widebar{F}} ]\Bigr] \\
    \nonumber &=& \langle \widebar{F} | V_{t_1 - t_0}[\Pi_F ] | \widebar{F} \rangle \langle F | V_{t_2 - t_1}[\Pi_{\widebar{F}} ] | F \rangle \\
    &=& P_{F \rightarrow \widebar{F}}(t_1 - t_0) P_{\widebar{F} \rightarrow F}(t_2 - t_1),
\end{eqnarray}
where the third row is obtained by taking into account that $\rho_{fd}(t) = 0$, and the fourth row follows from orthogonality of spaces $\mathbf{H}_f$ and $\mathbf{H}_d$. Finally, the probability $P(\neg F,\neg F)$ follows straightforwardly from the normalization condition $\sum_{O_i, O_j \in \{F, \neg F\}} P(O_i, O_j) = 1$, i.e.,
\begin{eqnarray}
    \nonumber P(\neg F, \neg F) &=& 1 - P_{F \rightarrow F}(t_1 - t_0) \\
    &-& P_{F \rightarrow \widebar{F}}(t_1 - t_0) P_{\widebar{F} \rightarrow F}(t_2 - t_1).
\end{eqnarray}


\bibliography{LibraryNeutrino}

\bibliographystyle{apsrev4-2}

\end{document}